\documentclass[10pt,twocolumn,prl,aps,amssymb,amsmath,tightenlines]{revtex4}

\usepackage{graphicx}
\usepackage{dsfont}
\newcommand{\re}{{\rm e}}
\newcommand{\ri}{{\rm i}}

\newcommand{\half}{\mbox{$\textstyle \frac{1}{2}$}}
\begin{document}
\title{Evolution speed of open quantum dynamics}
\author{Dorje C Brody$^1$ and Bradley Longstaff$^{2}$}

\affiliation{
$^1$Department of Mathematics, University of Surrey, Guildford GU2 7XH, UK \\ 
$^{2}$Department of Mathematics, Imperial College London, London SW7 2AZ, UK}

\begin{abstract}
The space of density matrices is embedded in a Euclidean space to deduce the dynamical equation satisfied by the state of an open quantum system. The Euclidean norm is used to obtain an explicit expression for the speed of the evolution of the state. The unitary contribution to the evolution speed is given by the modified skew information of the Hamiltonian, while the radial component of the evolution speed, connected to the rate at which the purity of the state changes, is shown to be determined by the modified skew information of the Lindblad operators. An open-system analogue of the quantum navigation problem is posed, and a perturbative analysis is presented to identify the amount of change on the speed. Properties of the evolution speed are examined further through example systems, showing that the evolution speed need not be a decreasing function of time. 
\end{abstract}

 
\maketitle

Understanding the speed of the evolution of a quantum state is of interest for a 
variety of reasons in quantum information science. As well as being of interest 
in its own right \cite{MT,Uf}, in implementing quantum 
algorithms for establishing communication and performing computation, for 
instance, the evolution speed determines how fast a given task can be processed. 
The speed also determines the sensitivity of quantum states against 
time evolution, and this information can be used to determine error bounds on 
quantum state estimation \cite{BH0}. 

In the case of a pure state undergoing a unitary time evolution, the evolution 
speed of the state in Hilbert space was identified by Anandan and Aharonov as 
twice the energy uncertainty of the system \cite{AA}. In the case of a mixed 
state, the speed of unitary time evolution in Hilbert space is somewhat reduced 
to twice the Wigner-Yanase skew information \cite{DCB0,Luo}. 
Interests in the evolution speed in the context of quantum-state estimation 
grew rapidly after the work in \cite{BC} that connected the estimation problem to 
the geometry of the quantum state space. By now there is a substantial body of 
literature that clarifies various aspects of the speed of unitary time evolution 
(see, e.g., \cite{Pati} and references cited therein). 

More recently, inspired in part by the desire to understand fundamental quantum 
limits to implementing quantum processes in more realistic environments, research 
activity into the study of evolution speed of open quantum systems has intensified 
(see \cite{DC} and references cited therein). 
In this connection it is worth noting that the notion of speed, which is the ratio 
of distance and time, crucially depends on the choice of the metric on the space of 
quantum states. For pure states, there is little ambiguity to the matter on account 
of the existence of a unique unitary-invariant Fubini-Study metric on the space of 
pure states \cite{DCB1}. However, for mixed-state density matrices, the structure 
of the state 
space is more intricate, and there is a range of different metrics one can impose, 
whose merits are dependant on the particular application one might consider. 
For instance, in \cite{AMR} the authors consider the parameter sensitivity of the 
state by examining the Fisher information associated with an open system 
dynamics. The trace norm of the difference of two density matrices is considered 
in \cite{Funo} to define distance, which is used to bound the 
evolution speed of an open-system dynamics. In \cite{Raam} the purity of the 
state is used to define distance, and an upper bound for the speed of evolution 
is obtained. In \cite{Deffner}, Uhlmann's fidelity \cite{U,Jozsa} between the initial 
state, assumed pure, and the terminal state is taken to define distance to obtain 
a bound on the evolution speed for general open dynamics. The fidelity-based 
measure is also considered in \cite{Taddei} where the minimum evolution time for 
general open-system dynamics is investigated, and shown that the closely-related 
work of \cite{Plenio} does not reproduce their results. This is natural because in 
\cite{Plenio} the relative purity is used to define distance, and hence one does 
not \textit{a priori} expect the results to coincide. In \cite{BG} the evolution speed 
of the state under the influence of non-Hermitian Hamiltonian is worked out. 

In the present paper we regard the density matrix as representing a vector 
in a Euclidean space, and take the natural Euclidean metric as defining the 
distance measure. This measure is distinct from the fidelity or relative purity 
measures, but arises naturally and 
exhibits similar characteristics. In this setup we consider a one-parameter family 
of states generated by a general open-system dynamics, and work out the 
explicit expression for the evolution speed, which consists of three terms: One 
associated with the unitary time evolution, given by the modified skew information 
of the Hamiltonian; one 
associated with the environmental influences generated by Lindblad 
operators; and one associated with the competition between the two. We also 
work out the radial component of the speed, which is shown to be related to the 
speed of the change of the purity ${\rm tr}({\hat\rho}^2)$ of the state. 
Surprisingly, this is determined by the modified skew information of 
the Lindblad operators. 
We then calculate the changes to the speed induced by a small perturbation 
of the Hamiltonian, when the ambient environmental influences, characterised 
by the Lindblad operators, cannot be controlled (an analogue of the Zermelo 
navigation problem \cite{BGM} for open quantum systems). The behaviours of 
the evolution speed are then studied in example systems: For a PT-symmetric 
quantum system we show that the existence of a phase transition leads to 
qualitatively different behaviours of the speed; while in a Bose-Hubbard 
system coupled to a reservoir we show that the evolution speed need not be 
decreasing in time, contrary to what other studies have suggested. 

We begin by remarking that the space of density matrices in a Hilbert space 
${\cal H}^n$ of dimension $n$ forms a subset of the interior of a sphere 
$S^{n^2-2}$ in a Euclidean space ${\mathds R}^{n^2-1}$ \cite{DCB1}. Thus every 
density matrix ${\hat\rho}$ can be thought of as being represented by a vector 
${\boldsymbol r} \in {\mathds R}^{n^2-1}$. There are various ways in which we 
can choose a system of coordinates for ${\mathds R}^{n^2-1}$, but here we 
consider the generalisation of the Bloch vector representation \cite{BK}. For this 
purpose we let $\{{\hat \sigma}_j\}_{j=0,\ldots,n^2-1}$ be an orthonormal basis for 
the linear space of bounded operators on ${\cal H}^n$  equipped with the 
Hilbert-Schmidt inner product 
$\langle {\hat\sigma},{\hat\tau}\rangle ={\rm tr}\left({\hat \sigma}^\dagger
{\hat \tau}\right)$. We set ${\hat \sigma}_0=n^{-1/2}{\mathds 1}$, hence the 
operators $\{{\hat \sigma}_j\}_{j=1,\ldots,n^2-1}$ are trace free, and together they 
satisfy the orthonormality condition $\langle {\hat \sigma}_i,{\hat \sigma}_j\rangle 
= \delta_{ij}$. For $n=2$ we may set $\{\hat\sigma_j\}_{j=1,\ldots,3}$ to $1/\sqrt{2}$ 
times the Pauli matrices; for $n=3$ we may set $\{\hat\sigma_j\}_{j=1,\ldots,8}$ to 
$1/\sqrt{2}$ times the 
Gell-Mann matrices, and so on. An arbitrary density matrix ${\hat\rho}$ 
can then be expressed in the form 
\begin{eqnarray}
{\hat\rho} = \frac{1}{\sqrt{n}}\, {\hat \sigma}_0 
+ \sum_{j=1}^{n^2-1} r_j \, {\hat \sigma}_j, 
\label{eq:1} 
\end{eqnarray}
where $r_j={\rm tr}({\hat\rho}{\hat \sigma}_j)$, $j=1,\ldots,n^2-1$, are the 
components of the vector ${\boldsymbol r} \in {\mathds R}^{n^2-1}$. For 
a pure state we have 
\begin{eqnarray} 
1 = {\rm tr}({\hat\rho}^2) = \frac{1}{n} + \sum_{j=1}^{n^2-1} r_j^2 ,
\end{eqnarray} 
from which it follows that the squared radius of the sphere $S^{n^2-2}$ in 
${\mathds R}^{n^2-1}$ is given by $1-n^{-1}$. 

We now consider a one-parameter family of density matrices ${\hat\rho}(t)$ 
parameterised by time $t$ that satisfy the dynamical equation 
\begin{eqnarray}
\label{eqn:liouvillian}
\partial_t {\hat\rho} = -\ri[{\hat H},{\hat \rho}] + 
\sum_k \left[ {\hat L}_k {\hat\rho} {\hat L}^\dag_k - \frac{1}{2} \left( 
{\hat L}^\dagger_k {\hat L}_k \rho + {\hat\rho} {\hat L}^\dagger_k {\hat L}_k 
\right) \right]
\label{eq:4} 
\end{eqnarray}
along with an initial condition ${\hat\rho}(0)$. The unitary part of the dynamics is 
described the Hamiltonian ${\hat H}$, and $\{{\hat L}_k\}$ is a family of Lindblad 
operators characterising the system interaction with its environment. 
Our strategy now is to identify the linear differential equation satisfied by the state 
${\boldsymbol r}$ that corresponds to the evolution equation (\ref{eq:4}) for the 
density matrix. For this purpose we substitute (\ref{eq:1}) in (\ref{eq:4}) to obtain 
\begin{eqnarray}
\sum_{j=1}^{n^2-1} {\dot r}_j {\hat\sigma}_j &=& -\ri \sum_{j=1}^{n^2-1} [{\hat H},
{\hat\sigma}_j] r_j + \frac{1}{n} \sum_k [ {\hat L}_k,{\hat L}_k^\dagger] 
\nonumber \\ & &  \hspace{-2.0cm} 
+ \sum_{j=1}^{n^2-1} \sum_k 
\left[ {\hat L}_k{\hat\sigma}_j{\hat L}_k^\dagger - \frac{1}{2} \left( 
{\hat L}^\dagger_k {\hat L}_k {\hat\sigma}_j + {\hat\sigma}_j {\hat L}^\dagger_k 
{\hat L}_k \right) \right] r_j . 
\label{eq:5} 
\end{eqnarray}
We multiply ${\hat\sigma}_i$ to both sides of (\ref{eq:5}) and take the trace, 
using the orthonormality relation $\langle {\hat\sigma}_i,{\hat\sigma}_j\rangle =
\delta_{ij}$, to deduce that $r_j$ satisfies the differential equation 
\begin{eqnarray}
{\dot r}_i = \sum_{j=1}^{n^2-1} \Lambda_{ij} r_j + b_i , 
\label{eq:6} 
\end{eqnarray}
where 
\begin{eqnarray}
\Lambda_{ij} &=& \, {\rm tr} \left[ -\ri\, [{\hat\sigma}_j,{\hat\sigma}_i] 
{\hat H} + \sum_k {\hat L}_k{\hat\sigma}_j{\hat L}_k^\dagger {\hat\sigma}_i
\right. \nonumber \\ && \qquad \left. - \frac{1}{2} \sum_k \left( 
{\hat L}^\dagger_k {\hat L}_k {\hat\sigma}_j {\hat\sigma}_i + 
{\hat L}^\dagger_k {\hat L}_k {\hat\sigma}_i {\hat\sigma}_j \right) \right] 
\end{eqnarray}
is a real matrix, and 
\begin{eqnarray}
b_i = \frac{1}{n} \sum_k {\rm tr} \left( [{\hat L}_k,{\hat L}^\dagger_k ] 
{\hat\sigma}_i \right) 
\end{eqnarray} 
is a real vector. If the Lindblad operators are Hermitian, or more 
generally if they are normal, then we have $b_j = 0$. There are also other 
circumstances in which ${\boldsymbol b}$ vanishes, for instance when 
there are two Lindblad operators given by ${\hat L}_1 = {\hat\sigma}_+$ and 
${\hat L}_2 = {\hat\sigma}_-$, where ${\hat\sigma}_\pm={\hat\sigma}_x\pm 
\ri{\hat\sigma}_y$.

From the linearity of the dynamics we can think of the right 
side of (\ref{eq:4}) as representing the action of a Liouville operator ${\cal L}$ 
(cf. \cite{wiener}) on ${\hat\rho}$, and write 
$\partial_t {\hat\rho}={\cal L}{\hat\rho}$ for (\ref{eq:4}), where  
${\hat\rho}$ is viewed as a vector on which the linear operator 
${\cal L}$ acts. The components ${\cal L}_{ij}$ of ${\cal L}$ in the basis 
$\{{\hat\sigma}_j\}$ are given by 
${\cal L}_{ij} =  \langle {\hat\sigma}_i, {\cal L}{\hat\sigma}_j\rangle  = 
{\rm tr}\left( {\hat\sigma}_i {\cal L}{\hat\sigma}_j \right)$, 
and we have 
$({\cal L}{\hat\rho})_j = \sum_i {\rm tr}( {\hat\sigma}_j {\cal L}{\hat\sigma}_i) 
\, {\rm tr} ( {\hat\sigma}_i {\hat\rho})$.
It follows that $\left({\cal L}{\hat\rho}\right)_0=0$, because 
${\rm tr}({\cal L}\hat \xi)=0$ for any $\hat \xi$.
Therefore, writing 
\begin{eqnarray}
{\dot r}_j = \sum_{i=1}^{n^2-1} {\cal L}_{ji} r_i + \frac{1}{\sqrt{n}}\,{\cal L}_{j0}
\end{eqnarray}
we deduce from (\ref{eq:6}) that the matrix elements of the Liouville operator are 
given by 
${\cal L}_{ji} = \Lambda_{ji}$ for $i,j\neq0$ and ${\cal L}_{j0} = \sqrt{n}\,b_j$. Because 
the real matrix ${\cal L}_{ji}$ is not symmetric, its eigenvalues are either real or 
else come in complex conjugate pairs, such that the real parts of the eigenvalues 
are nonpositive, thus generating a completely positive map on the space of 
density matrices \cite{Baum08}. 

Having obtained the dynamical equation satisfied by the state vector 
${\boldsymbol r} \in {\mathds R}^{n^2-1}$ we are now in the position to determine 
the squared speed of evolution:
\begin{eqnarray}
\label{eqn:speed sq}
v^2(t) = \sum_{j=1}^{n^2-1} {\dot r}_j^2 = {\rm tr} \left[ ({\cal L}{\hat\rho})^2\right]. 
\end{eqnarray}
To proceed let us write the time derivative of the state ${\hat\rho}$ in the form 
$\mathcal{L}{\hat\rho} = -\ri [{\hat H},{\hat\rho}] + \mathcal{D}{\hat\rho}$,
thus isolating the dissipator term 
\begin{eqnarray}
\label{eqn:dissipator}
\mathcal{D}{\hat\rho} = \sum_k \left[ {\hat L}_k {\hat\rho} {\hat L}^\dag_k - \frac{1}{2} 
\left( {\hat L}^\dagger_k {\hat L}_k \rho + {\hat\rho} {\hat L}^\dagger_k {\hat L}_k 
\right) \right]
\end{eqnarray}
from the unitary part of ${\cal L}$. Squaring $\mathcal{L}{\hat\rho}$ and taking 
the trace, we obtain: 
\begin{eqnarray}
v^2(t) &=& 2\left[{\rm tr}\left({\hat H}^2{\hat\rho}^2\right) - {\rm tr}\left( {\hat H} 
{\hat\rho} {\hat H} {\hat\rho}\right)\right] \nonumber \\ && \quad  
- 2\ri\, {\rm tr}\left( {\hat\rho}\, [{\cal D}{\hat\rho},{\hat H}]\right) + {\rm tr} \left[\left(\mathcal{D}{\hat\rho}\right)^2\right].
\label{eq:15}
\end{eqnarray}
There are three terms contributing to the speed of evolution; the first arising purely 
from the unitary evolution and the third arising purely from the dissipator term, while 
the second term represents in some sense the competition between the Hamiltonian 
and the Lindblad operators. To see this we note that the cross term in (\ref{eq:15}) 
can alternatively be written in the form 
\begin{eqnarray}
-2\ri \sum_k \left[ 
{\rm tr} \left( {\hat\rho}\, [{\hat L}_k {\hat\rho} {\hat L}^\dagger_k,{\hat H}]\right) 
+ \half {\rm tr} \left({\hat\rho}^2 [{\hat H},{\hat L}^\dagger_k {\hat L}_k] \right) \right] , 
\nonumber
\end{eqnarray}
which vanishes if ${\hat L}_k^\dagger={\hat L}_k$ and $[{\hat L}_k,{\hat H}]=0$, thus 
representing the lack of compatibility between ${\hat H}$ and $\{{\hat L}_k\}$.

The contribution to the squared speed from the unitary evolution (the first term
in (\ref{eq:15})) has recently been identified in \cite{CPBM} as the speed of 
the evolution of the state, when the distance is measured with respect to the 
Euclidean angular separation. This term resembles, but is different from, the 
Wigner-Yanase skew information $I = {\rm tr}( {\hat H}^2{\hat\rho}) - {\rm tr}({\hat H}
{\sqrt{\hat\rho}}{\hat H}{\sqrt{\hat\rho}})$. The apparent discrepancy between this 
result and that obtained in \cite{DCB0} is that here we measure the speed with 
respect to the Euclidean norm in ${\mathds R}^{n^2-1}$, whereas in \cite{DCB0} 
the evolution speed of the state is obtained using the Hilbert space norm. While 
the latter is more useful in the context of state estimation (because the Fisher-Rao 
metric for unitary evolution is given by the skew information), as remarked in 
\cite{CPBM} for the analysis of the evolution speed and time, the use of the 
Euclidean metric is computationally more effective for it does not involve taking 
the square-root of the density matrix. We shall refer to $S(X)={\rm tr}(
{\hat X}^\dagger{\hat X}{\hat\rho}^2) - {\rm tr}({\hat X}{\hat\rho}{\hat X}^\dagger
{\hat\rho})$ as the `modified skew information' for the operator ${\hat X}$. 

In contrast to unitary time evolution, in an open system the velocity will in general 
obtain a radial component so that the purity ${\rm tr}({\hat\rho}^2)$ changes. To 
see this, consider the squared magnitude of the radial velocity
$v^2_R(t) = ({\boldsymbol r}\cdot {\dot{\boldsymbol r}})^2/
({\boldsymbol r}\cdot {\boldsymbol r}) = [{\rm tr}({\hat\rho} \mathcal{L}
{\hat\rho})]^2/{\rm tr}[({\hat\rho}-n^{-1}{\mathds 1})^2]$, 
which vanishes for unitary dynamics. In \cite{Raam} an upper bound 
for the numerator term $[{\rm tr}({\hat\rho} \mathcal{L}{\hat\rho})]^2$
is obtained using 
Cauchy-Schwarz inequality, but in fact a short calculation shows remarkably that  
\begin{eqnarray}
v_R(t) = \sum_k \frac{S(L_k)}{\sqrt{{\rm tr} \left[({\hat\rho}-n^{-1}{\mathds 1})^2\right]}} . 
 \label{eq:18}
\end{eqnarray}
In other words, the speed of the change of the purity is given by twice the modified 
skew information associated with the Lindblad operators. 

Let us turn to consider optimisation. Two scenarios that might arise in the context of 
open quantum systems are: (i) to maximise the speed over all Liouville 
operators; and (ii) to maximise the speed over all Hamiltonians for a fixed 
open environment $\{{\hat L}_k\}$. The latter problem arises when an 
experimentalist has no control over the environmental influences, but nonetheless 
can set the Hamiltonian so as to implement a rapid state transportation (an 
open-system analogue of the quantum navigation problem \cite{BGM}). Under a 
unitary evolution, the solution to problem (i) is obtained by maximising the modified 
skew information. 
For an open system, finding general solutions to these problems is 
nontrivial, in part because of the competition between ${\hat H}$ and $\{{\hat L}_k\}$, 
i.e. the second term in (\ref{eq:15}) can \textit{a priori} be positive or negative. We 
can nevertheless explore a perturbative analysis. Specifically, for problem (ii) we 
perturb the Hamiltonian ${\hat H}$ in the direction of ${\hat\Delta}$ by a small 
amount $\epsilon$, i.e. we let ${\hat H}\to{\hat H}+\epsilon{\hat\Delta}$, and work 
out how much the speed changes in the limit $\epsilon\to0$. A calculation shows 
that this is given by 
\begin{eqnarray}
\delta v^2 &=& 2\left[{\rm tr}\left(({\hat H}{\hat\Delta}+{\hat\Delta}{\hat H}){\hat\rho}^2\right) 
- 2 {\rm tr}\left( {\hat H} {\hat\rho} {\hat\Delta} {\hat\rho}\right)\right] \nonumber \\ && 
\hspace{-1.350cm} -2\ri \sum_k \left[ {\rm tr} \left( {\hat\rho}\, [{\hat L}_k {\hat\rho} 
{\hat L}^\dagger_k,{\hat\Delta}]\right) 
+ \half {\rm tr} \left({\hat\rho}^2 [{\hat\Delta},{\hat L}^\dagger_k {\hat L}_k] \right) \right] ,
\label{eq:x13} 
\end{eqnarray}
which can be used, e.g., to numerically explore the optimal way to modify the Hamiltonian 
under uncontrollable environmental influences. 

We now examine the behaviour of the evolution speed via illustrative 
examples. For the first example we take the Hamiltonian to be ${\hat H}=
\frac{1}{2}g {\hat\sigma}_z$ and the Lindblad operator to be ${\hat L}=
\sqrt{\gamma}{\hat\sigma}_z$, thus describing pure dephasing of the two-level 
system with a decay rate $\gamma$. In this example we find that 
$v^2(t) = \re^{-4\gamma t}(4\gamma^2 + g^2)[r^2_x(0) + r^2_y(0)]$, and hence 
that $v(t)$ decreases exponentially in time. 

As a variant of the previous example, suppose that the Hamiltonian is given by 
${\hat H}=\frac{1}{2}g{\hat\sigma}_x$ so that it no longer commutes with the 
Lindblad operator ${\hat L}=\sqrt{\gamma}{\hat\sigma}_z$. This is perhaps the 
simplest example of a passive PT-symmetric quantum system. In this case we 
have 
\begin{eqnarray}
\Lambda = \begin{pmatrix} -2\gamma & 0 & 0\\0 &-2\gamma & -g\\0 
& g & 0 \end{pmatrix} ,
\end{eqnarray}
and the four eigenvalues of the Liouville operator are thus given by 
$0$, $-2\gamma$, and $-\gamma \pm \sqrt{\gamma^2-g^2}$.
We see that in the region of unbroken PT-symmetry where $g<\gamma$, 
all the eigenvalues are real; at the exceptional point $g=\gamma$ the PT-symmetry 
gets broken and in the symmetry-broken phase where $g>\gamma$ the eigenvalues 
are either all real or else come in complex conjugate pairs. We expect to observe 
different behaviours of the system in each of these phases.
Indeed, the solutions to (\ref{eq:6}) are given by: $r_x(t)=\re^{-2\gamma t} r_x(0)$, 
$r_y(t) = \re^{-\gamma t}[\left(\cos \omega t - (\gamma/\omega) \sin\omega t\right)r_y(0) 
- (g/\omega) \sin \omega t \, r_z(0)]$ and $r_z(t) =  \re^{-\gamma t}[ (g/\omega) 
\sin \omega t\, r_y(0) + \left(\cos \omega t + 
(\gamma/\omega)\sin \omega t\right)r_z(0)]$, with $\omega = \sqrt{g^2-\gamma^2}$.
They are oscillatory in the broken phase $g>\gamma$, whereas in the 
unbroken phase $g<\gamma$ the oscillations associated with the unitary part are 
completely suppressed. 

\begin{figure}[h]
      \centering
        \includegraphics[width=0.24\textwidth]{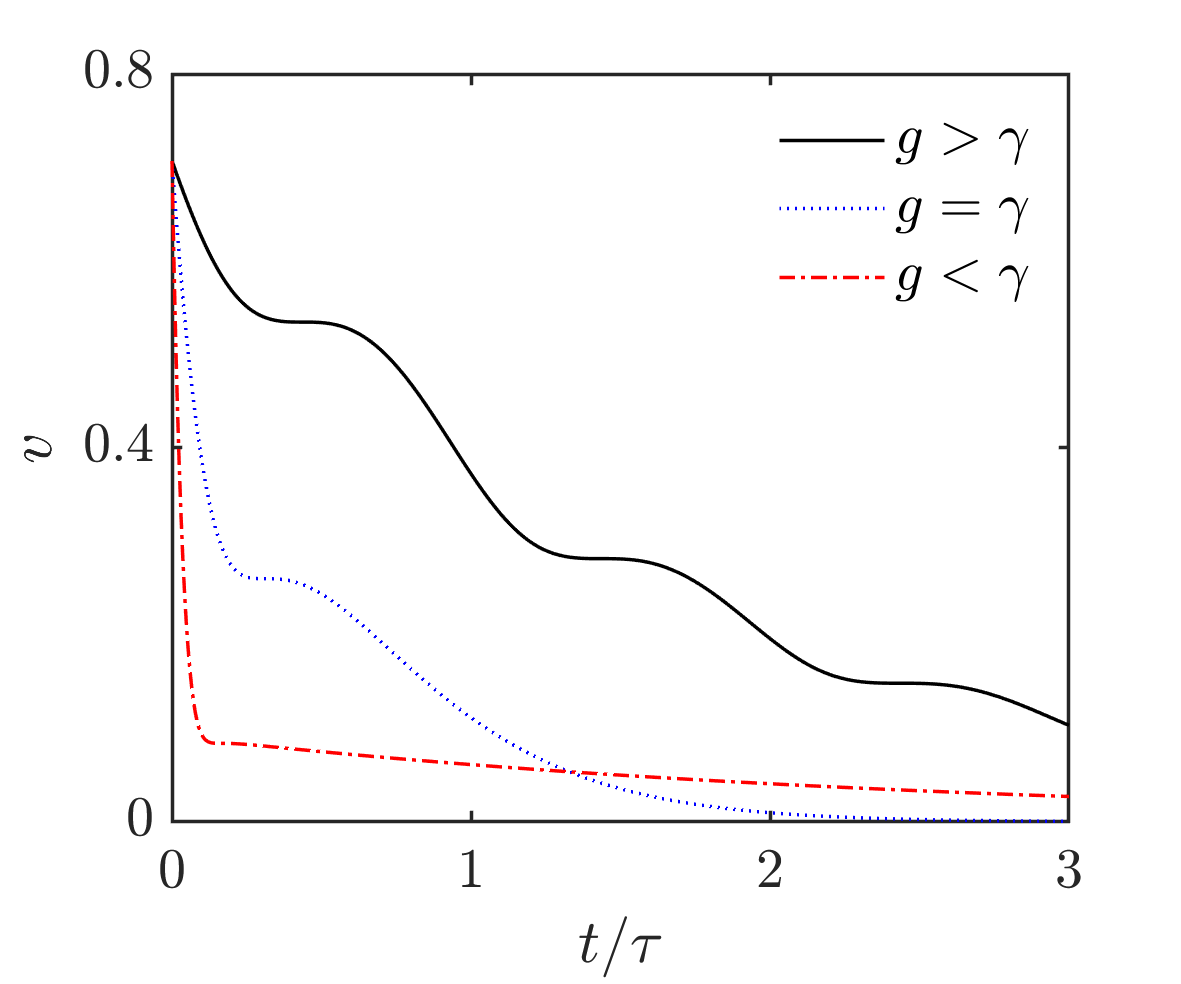}\hfill
        \includegraphics[width=0.24\textwidth]{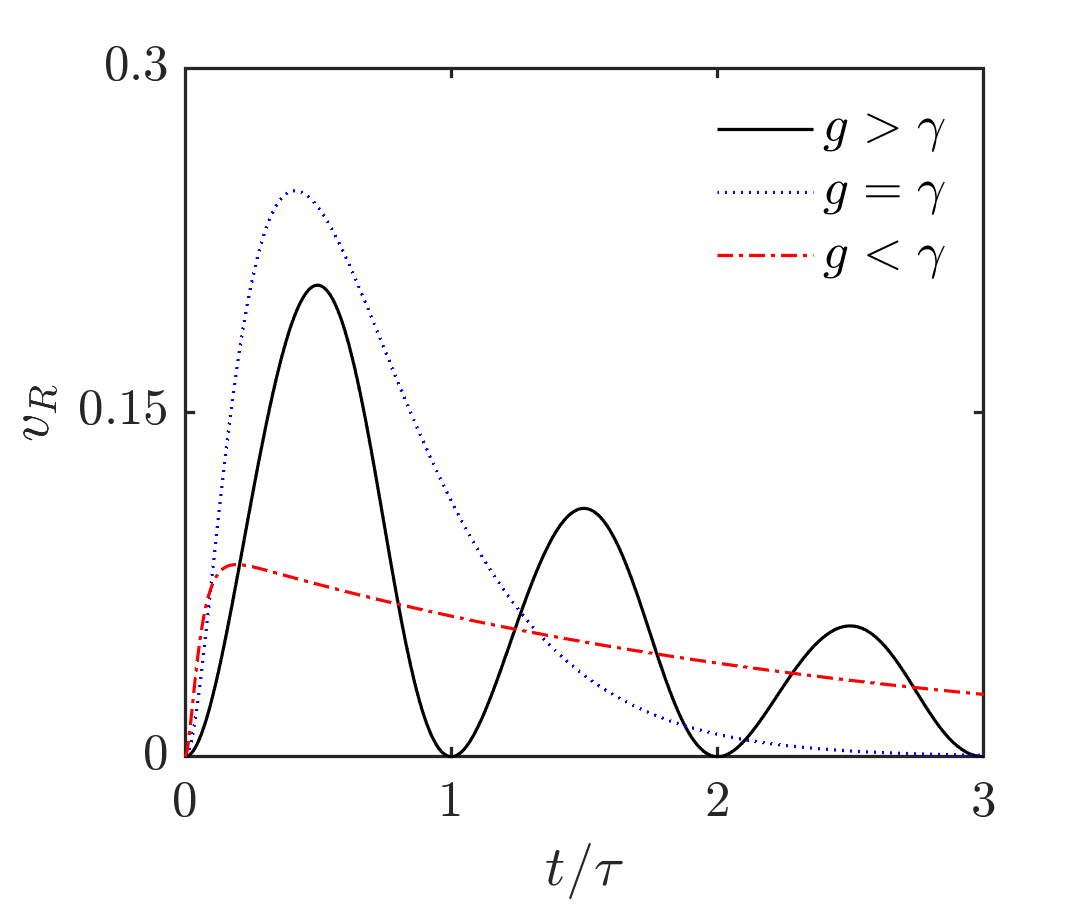}\hfill
        \includegraphics[width=0.24\textwidth]{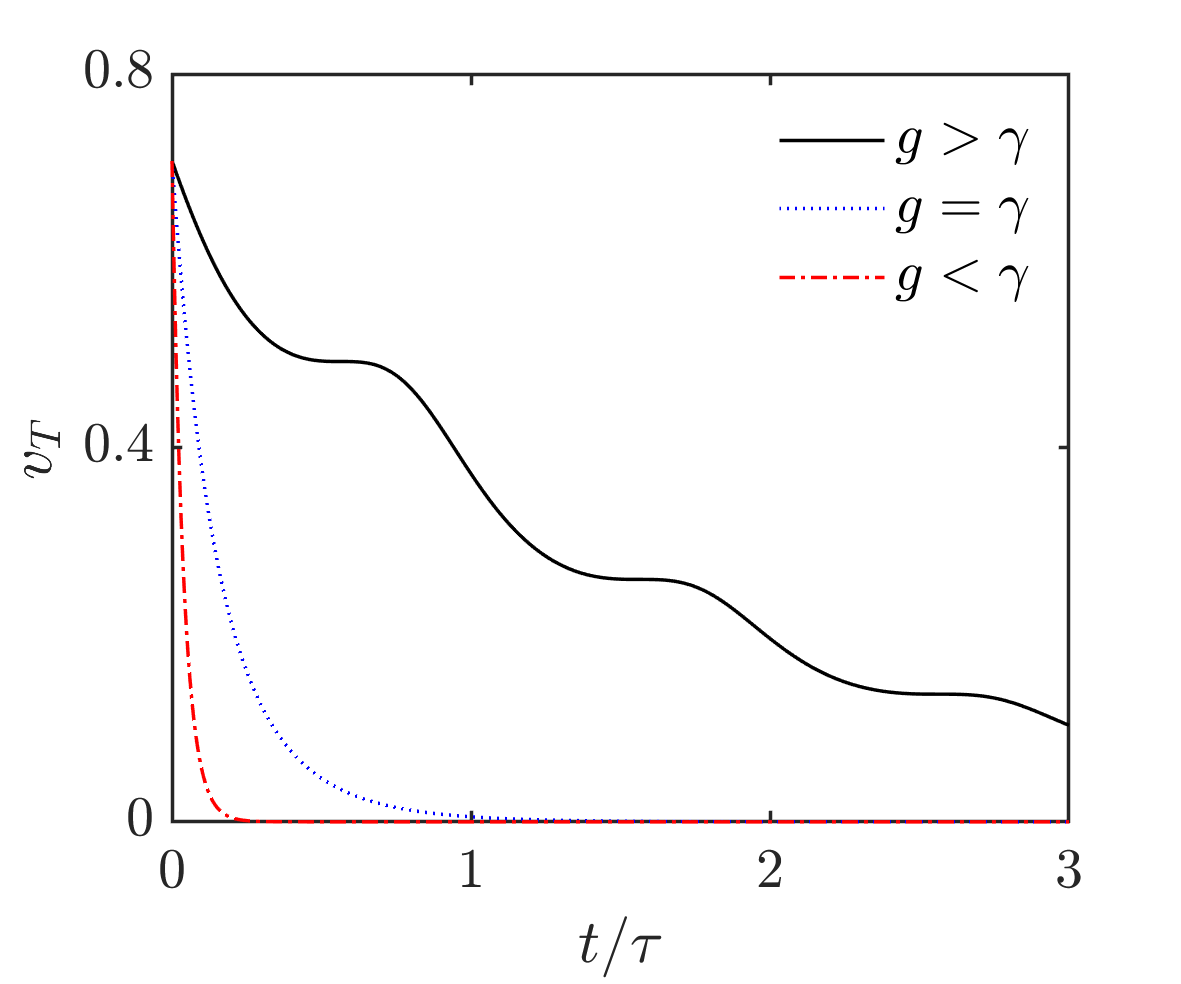}\hfill
        \includegraphics[width=0.24\textwidth]{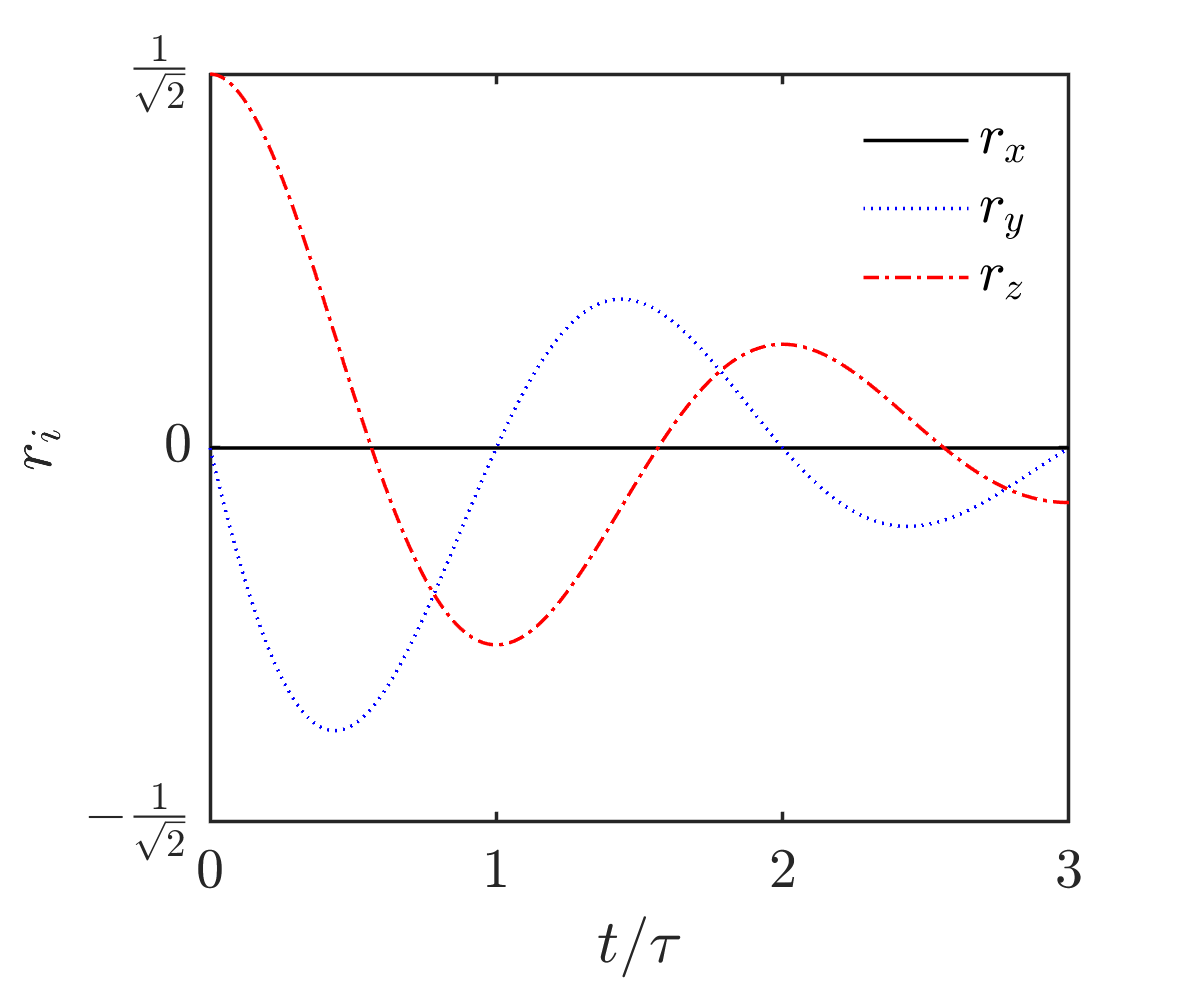}\hfill
\caption{\textit{Evolution speed for a PT-symmetric quantum system}. 
The behaviour of the evolution speed $v$ as a function of time is shown here 
(top left) when the system is initialised in the spin-$z$ up state. The 
radial $v_R$ (top right) and tangential $v_T$ (bottom left) components are 
also shown. Results for the decay rates $\gamma = 0.2, 1, 4$, corresponding to 
the broken, critical and unbroken PT-symmetry phases, are shown in each plot 
when $g=1$. The components of the Bloch vector in the broken phase are pictured 
on the bottom right. Time is measured in units of the period 
$\tau=\pi/\sqrt{g^2-\gamma^2}$ with $\gamma = 0.2$.}
\label{fig:pt deph} 
\end{figure}

The speed, and the corresponding radial component, are obtained by inserting these 
expressions into equations (\ref{eqn:speed sq}) and (\ref{eq:18}) respectively. From 
these, the squared magnitude of the tangential component $v^2_T(t) = v^2(t) - v^2_R(t)$ 
of the velocity can be determined. 
The different components of the speed are shown in Figure \ref {fig:pt deph} for a 
system prepared in the spin-$z$ up state $|\psi(0)\rangle  = |\!\!\uparrow\rangle$. 
Because this is an eigenstate of ${\hat L}$, we have $v_R(t)=0$ at $t=0$. The 
behaviour of the speed varies between the broken and unbroken PT phases. In the 
broken phase, the speed exhibits a decay superimposed with oscillations. Here $v_R(t)$ performs periodic 
oscillations with the period $\tau=\pi/\sqrt{g^2-\gamma^2}$, where the minima 
correspond to the times at which the Bloch vector is aligned with the $z$-axis, i.e. 
when $\hat\rho$ is an eigenstate of ${\cal D}$. 
Moving into the unbroken 
phase, the speed decays rapidly at short times and the oscillation in 
$v_T(t)$ is completely damped out. However, in this phase 
the velocity remains nonzero for a longer duration, with a small nonzero radial 
component remaining once the tangential component has vanished. 

In the previous examples, as well as a number of other examples we considered, 
the evolution speed is a decreasing function of time. Indeed, in \cite{Plenio} 
the Cauchy-Schwarz inequality is applied to obtain a bound on the speed of relative 
purity change, which shows that this speed is decreasing in time, and one might 
conjecture that the evolution speed of the state is therefore also decreasing in time. 
However, this is not the case in general. We shall demonstrate this by means of a 
counterexample based on a driven dissipative Bose-Einstein condensate (BEC) in 
a two-site optical lattice, previously studied in the context of dissipative state 
preparation \cite{Diehl}. The unitary dynamics are generated by the Bose-Hubbard 
Hamiltonian
$\hat H = -\Omega \hat{J}_x + U \hat{J}^2_z$,
where $\Omega$ is the coupling strength between the two sites, $U\geq0$ is the 
repulsive on-site interaction strength and the angular momentum operators 
$\hat{J}_i$ satisfy the $\mathfrak{su}(2)$ commutation relations 
$[\hat{J}_i,\hat{J}_j] = \ri \varepsilon_{ijk}\hat{J}_k$. Coupling the bosons on the 
lattice to a reservoir leads to dissipation that can be described by the Lindblad 
operator $\hat L = \sqrt{\gamma} (\hat{J}_z - \ri\hat{J}_y)$. As the number operator 
$\hat N$ commutes with each $\hat{J}_i$, the particle number $N$ is conserved 
and we may thus restrict the analysis to the Hilbert subspace ${\cal H}^{N+1}$ of 
fixed particle number.

\begin{figure}[h]
      \centering
        \includegraphics[width=0.24\textwidth]{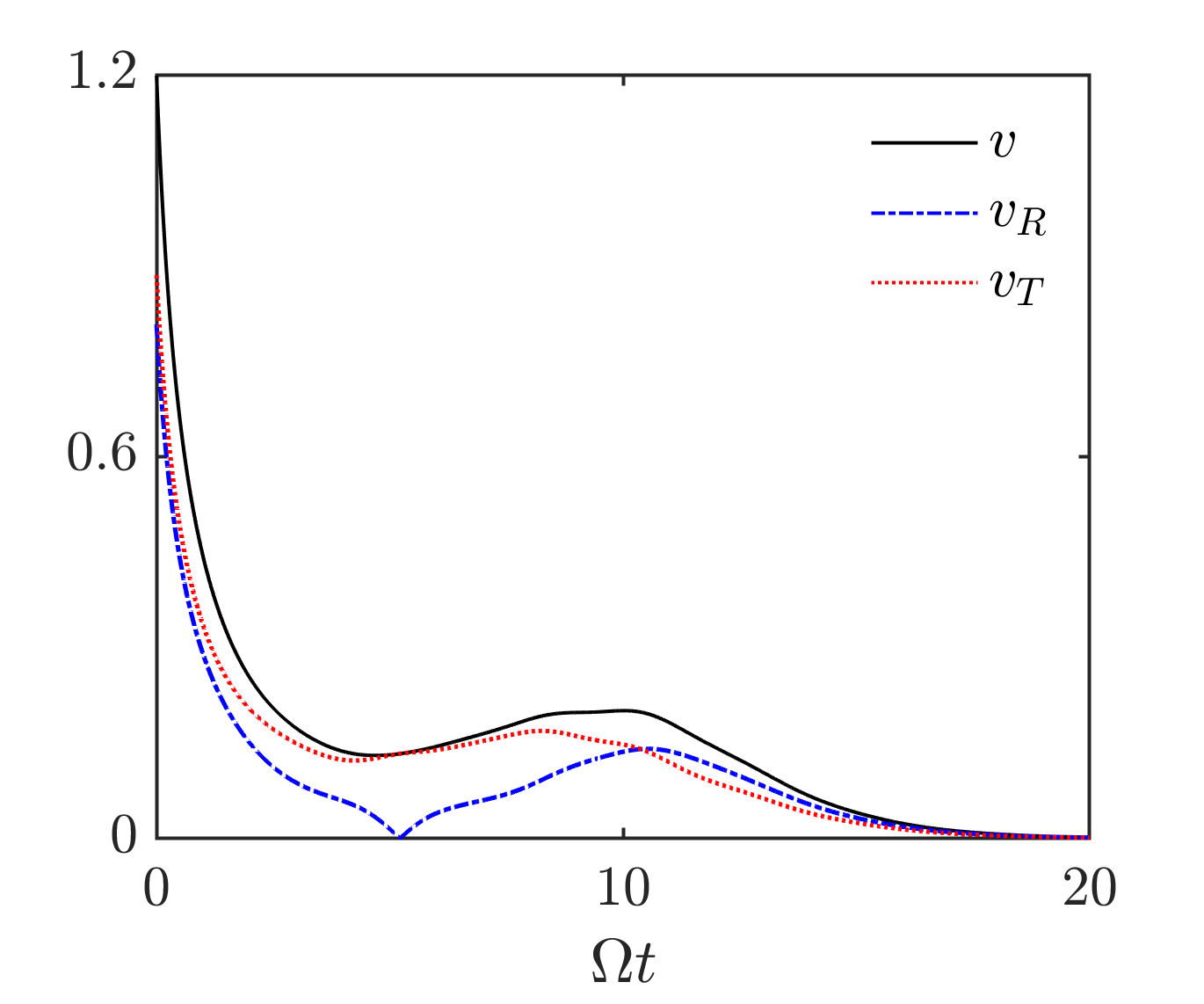}\hfill
        \includegraphics[width=0.24\textwidth]{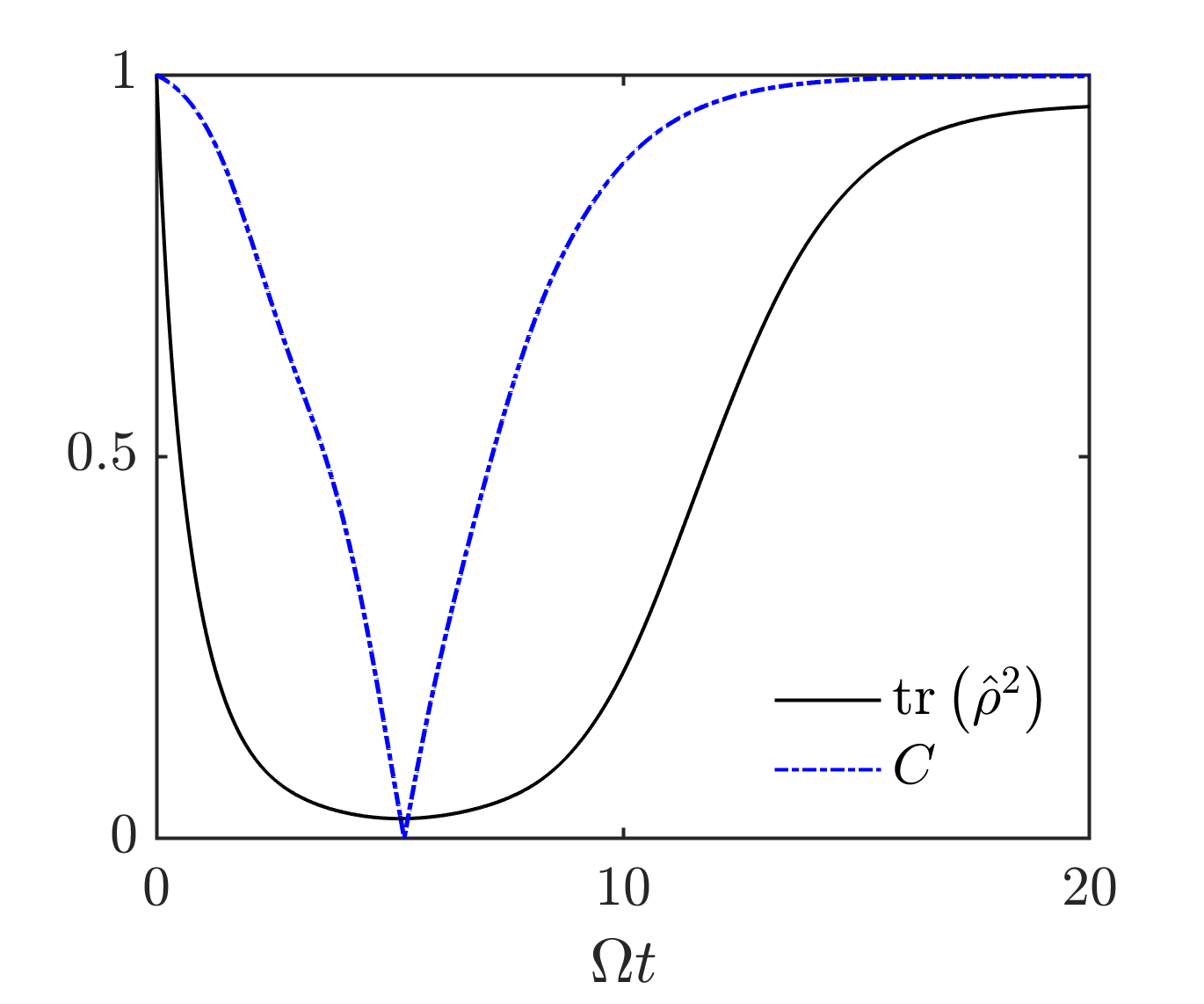}\hfill
        \includegraphics[width=0.24\textwidth]{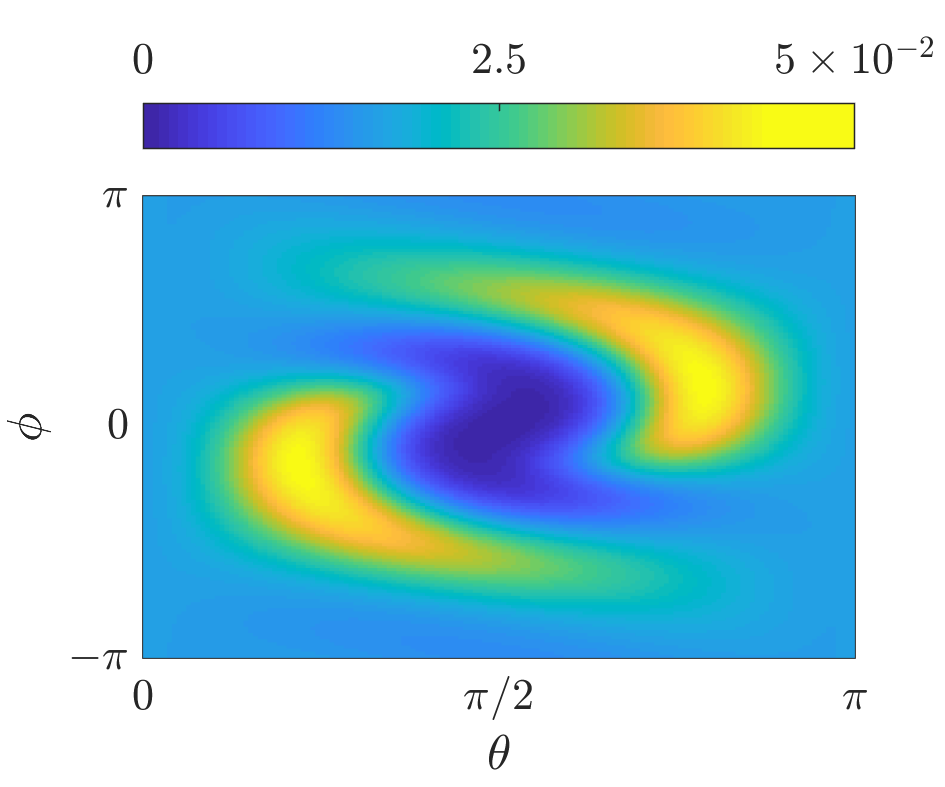}\hfill
        \includegraphics[width=0.24\textwidth]{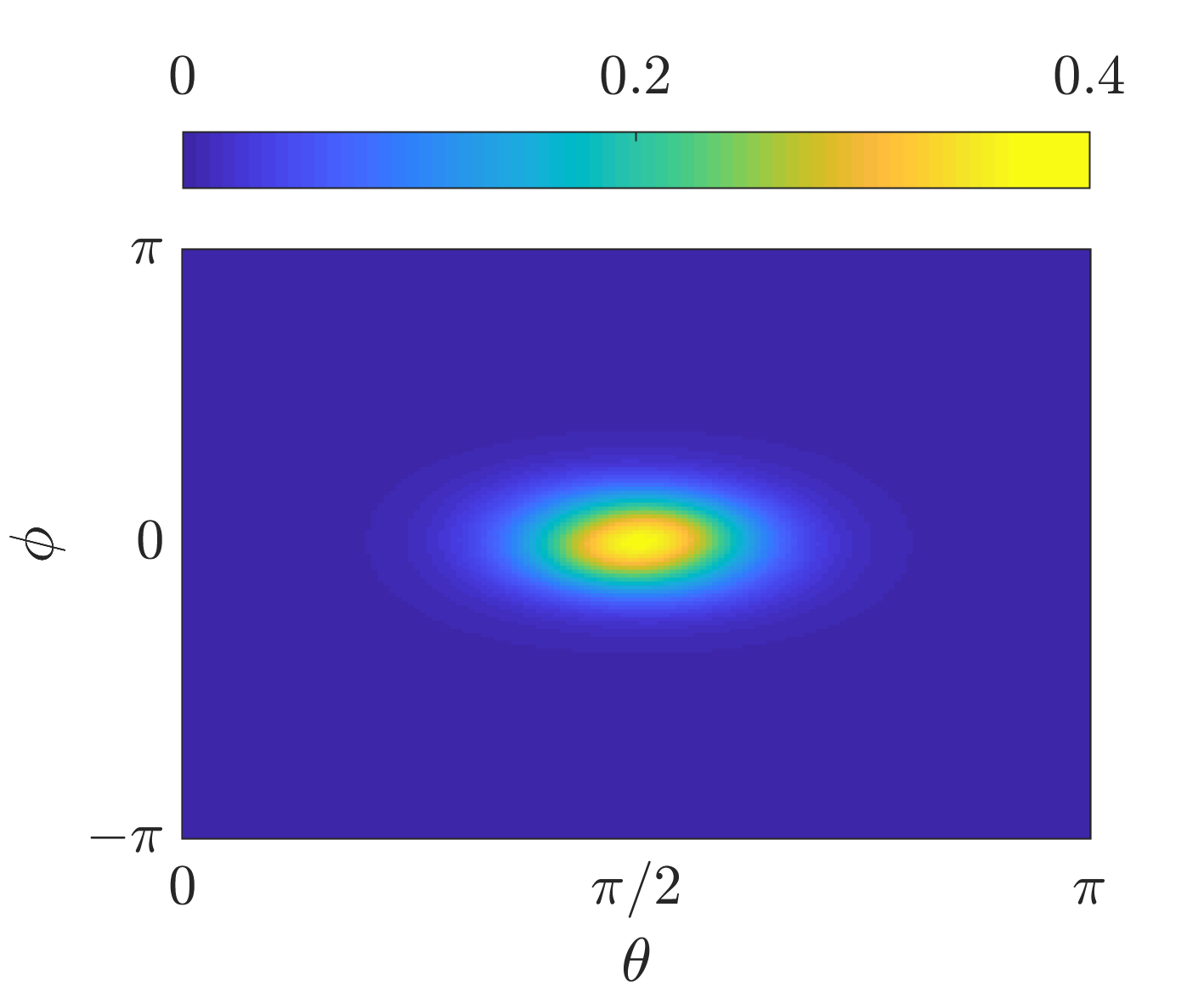}\hfill
\caption{\textit{Evolution speed in an open BEC system}. 
The evolution speed $v$, together with the radial $v_R$ and tangential 
$v_T$ components, is shown (top left) for an initial BEC state with the 
parameters $\theta=\pi/2, \phi=\pi$ and $N=50$. The purity ${\rm tr}({\hat\rho}^2)$ 
of the state and the phase coherence $C=2|\langle\hat{J}_x + {\rm i} \hat{J}_y
\rangle|(N^2-4\langle \hat{J}_z\rangle^2)^{-1/2}$ between the two sites are shown 
(top right). The Husimi function of the state at $\Omega t\approx 5$ and 
$\Omega t \approx 10$ is depicted on the bottom left and right, respectively. In 
each plot time is measured in units of the inverse tunneling rate, 
$UN = 0.8\Omega$ and $\gamma N = 0.8\Omega$.}
\label{fig:bh plots} 
\end{figure}

We worked out the evolution speed of an initial pure BEC state, 
represented by the $SU(2)$ coherent state 
$|\theta,\phi\rangle = \exp[\ri\theta(\hat{J}_x \sin \phi - \hat{J}_y \cos \phi)]|N,0\rangle$, 
where the Fock state $|N,0\rangle$ corresponds to all $N$ particles in the first lattice 
site. 
The result, plotted in Figure~\ref{fig:bh plots}, shows that the speed of evolution can 
increase. Initially, the Husimi (Fushimi) function 
$Q(\theta,\phi) = \langle \theta, \phi|\hat\rho |\theta,\phi\rangle$ of the state rapidly 
delocalises in phase space (lower left panel) and the speed slows down. However, the 
distribution then spirals towards the origin (a sink in the semiclassical limit), and, 
depending on the parameter choice, at 
the beginning of this localisation the speed can temporarily increase. As the state 
tends towards the steady state the speed then decreases again to zero. 
The initial loss of phase coherence between the two sites (upper right panel) 
indicates destruction of the 
condensate. The point at which the phase coherence is completely lost is the point at 
which the radial component of the velocity goes to zero, and the point at which the 
state is most mixed.

In summary, we have derived a closed-form expression for the evolution speed $v(t)$ 
of the state, which shows that it consists of three terms corresponding to the unitary 
contribution, the Lindblad contribution, and the competition of the two. We have also 
worked out the radial component, connected to the purity change, and showed that 
this is given by the modified skew information for the Lindblad operators. We examined 
example systems that show that the speed of evolution is typically decreasing in time, 
but this need not be the case in general. Our results on the evolution speed open up a 
new challenge of maximising $v(t)$ over all Liouville operators ${\cal L}$, as well as 
solving the open-system quantum navigation problem. \\ 

\textit{Note added}. While completing this work we came across a closely-related 
work \cite{CPBM}, in which the Euclidean norm is used to investigate bounds on the 
evolution time for general open systems. Various merits in the use of the Euclidean 
norm are also discussed therein. 


\end{document}